\documentclass[10pt,reqno,final]{amsart}
\usepackage{amsfonts} 
\usepackage{amsthm}
\usepackage{amsmath}
\usepackage{amssymb}
\usepackage{mathrsfs}
\usepackage[latin1]{inputenc}
\usepackage{color}
\usepackage{graphicx}
\usepackage{enumerate}
\usepackage{bbm}
\usepackage{changes} 
\newcommand*{\C}{\mathbb{C}}

\newcommand*{\N}{\mathbb{N}}

\newcommand*{\R}{\mathbb{R}}

\numberwithin{equation}{section}
\newtheorem{satz}{Satz}[section]

\newtheorem{theorem}[satz]{Theorem}

\newtheorem{lemma}[satz]{Lemma}
\newtheorem{remark}[satz]{Remark}
\newtheorem{definition}[satz]{Definition}

\bibliography{bibfile}
\begin{document}
\title[Stochastic Quantization]{Stochastic Quantization for the fractional Edwards Measure.}
\author[Wolfgang Bock]{Wolfgang Bock}
\address{Technomathematics Group, University of Kaiserslautern}
\email{bock@mathematik.uni-kl.de}

\author[Torben Fattler]{Torben Fattler}
\address{Functional Analysis and Stochastic Analysis Group, University of Kaiserslautern}
\email{fattler@mathematik.uni-kl.de}

\author[Ludwig Streit]{Ludwig Steit}
\address{BiBoS, Universität Bielefeld, Germany;
	CCM, Unversidade da Madeira, Funchal, Portugal and
	Physics Deptartment, MSU-IIT, Iligan, Philipinnes }
\email{streit@uma.pt}
\thanks{}
\date{\today}
\subjclass{} %
\keywords {}%
%
%
\begin{abstract}
We prove the existence of a diffusion process whose invariant measure is the fractional polymer or Edwards measure for fractional Brownian motion in dimension $d\in\mathbb{N}$ with Hurst parameter $H\in(0,1)$ fulfilling $dH < 1$.
The diffusion is constructed via Dirichlet form techniques in infinite dimensional (Gaussian) analysis. Moreover, we show that the process is invariant under time translations. 
\end{abstract}
\maketitle
\section{Introduction}
For a given probability measure $\nu$ on a measureable space $X$ the stochastic quantization of $\nu$ means the construction of a Markov process which has $\nu$ as an invariant measure. Stochastic quantization has been studied first by Parisi and Wu for applications in quantum field theory, which were extended to Euclidean quantum fields. 

The two-dimensional polymer measure is informally given as
$$
d\mu_g = Z^{-1} e^{-gL} d\mu_0,$$
where $\mu_0$ denotes the Wiener measure, $L$ the self-intersection local time of Brownian motion and $Z$ is a normalization constant. In the two-dimensional case stochastic quantization for this "polymer measure" has been studied by Albeverio, Roeckner, Hu and Zhou \cite{ARHZ}. Here we follow their approach, using Dirichlet forms also in the fractional case. 

Intersection local times $L$ of Brownian motion have been studied for a long
time and by many authors, see e.g.~\cite{ARHZ},\cite{bass}, \cite{fcs}-\cite{dvor2}, \cite{he}, \cite{imke}, \cite{legall}, \cite{lyons} and \cite{sym}-\cite{yor2}, 
the intersections of Brownian motion paths have been studied even since the
Forties \cite{levy}. One can consider intersections of sample paths with
themselves or e.g. with other, independent Brownian motions e.g. \cite{wolp}, one
can study simple \cite{dvor2} or $n$-fold intersections e.g. \cite{dvor3}, \cite{lyons} and one
can ask all of these questions for linear, planar, spatial or - in general -
$d$-dimensional Brownian motion: self-intersections become
increasingly scarce as the dimension $d$ increases.

A somewhat informal but very suggestive definition of self-intersection
local time of 
a Gaussian process $Y$
is in terms of an integral over Dirac's -
or Donsker's - $\delta $-function
\[
L(Y)\equiv \int d^2t\,\delta (Y(t_2)-Y(t_1)), 
\]
where for now $Y=B$ is a Brownian motion, intended to sum up the contributions from each pair of ''times'' $t_1,t_2$
for which the process $Y$ is at the same point. In Edwards' modeling
of long polymer molecules by Brownian motion paths, $L$ is used to model the
''excluded volume'' effect: different parts of the molecule should not be
located at the same point in space. As another application, Symanzik \cite
{sym} introduced $L$ as a tool in constructive quantum field theory.

A rigorous definition, such as e.g. through a sequence of Gaussians
approximating the $\delta $-function, will lead to increasingly singular
objects and will necessitate various ''renormalizations'' as the dimension d
increases. For $d>1$ the expectation will diverge in the limit and must be
subtracted \cite{legall}, \cite{varadhan}, as a side effect such a local time will then no
more be positive. For $d>3$ various further renormalizations have been
proposed \cite{watanabe} that will make $L$ into a well-defined generalized
function of Brownian motion. For $d=3$ a multiplicative renormalization
gives rise to an independent Brownian motion as the weak limit of
regularized and subtracted approximations to $L$ \cite{yor2}; another
renormalization has been constructed by Westwater to make the Gibbs factor $%
e^{-g\cdot L}$ of the polymer model well-defined \cite{west}.

In this article we first introduce the setting along the lines of white noise or Gaussian analysis, using  the fractional white noise measure. This can be compared to the approach in \cite{OH03}. Moreover the results from \cite{Po97} concerning the gradient are extended to this setting. 
We show in the framework of Dirichlet forms, that there exists a Markov process which has the fractional polymer measure as invariant measure. The proof is based on the results of \cite{Hu2001} and \cite{HNS06}, which show that the self-intersection local time in the case $Hd<1$ is Meyer-Watanabe differentiable. The closability of the gradient Dirichlet form is then shown by an integration by parts argument. The irreducibility follows as in the Brownian case, see \cite{ARHZ}.

\section{Framework}
For $d\in\mathbb{N}$ and \emph{Hurst parameter} $H\in(0,1)$ \emph{fractional Brownian motion in dimension $d$} is a $\mathbb{R}^d$-valued centered Gaussian process $\big(B^{\scriptscriptstyle{H}}_t\big)_{t\ge 0}$
with covariance
\begin{align*}
\text{cov}_{\scriptscriptstyle{H}}(t,s):=\mathbb{E}\big[B^{\scriptscriptstyle{H}}_t B^{\scriptscriptstyle{H}}_s\big]=\frac{1}{2}\left(
t^{2H}+s^{2H}-|t-s|^{2H}\right),\quad s,t\in [0,\infty). 
\end{align*}
For $s\in(0,\infty)$ let $\Theta_s:=\mathbbm{1}_{[0,s)}$ and set $\big(\Theta_{s},\Theta_{t}\big)_{\scriptscriptstyle{H}}:=\text{cov}_{\scriptscriptstyle{H}}(t,s)$ for $s,t\in [0,\infty)$. Moreover, let $X:=\text{span}\big\{\Theta_s\,\big|\,s>0\big\}$. Hence $x,y\in X$ are of the form
\begin{align*}
x=\sum_{i=1}^n\alpha_i\Theta_{s_{i}},\quad y=\sum_{j=1}^m\beta_j\Theta_{t_{j}}
\end{align*}
with $n,m\in\mathbb{N}$ and
\begin{align*}
\big(x,y\big)_{\scriptscriptstyle{H}}:=\sum_{i=1}^n\sum_{j=1}^m\alpha_i\,\beta_j\big(\Theta_{s_i},\Theta_{t_j}\big)_{\scriptscriptstyle{H}}
\end{align*}
defines an inner product on $X$. Taking the abstract completion of the inner product space $\big(X,(\cdot,\cdot)_{\scriptscriptstyle{H}}\big)$ we obtain a Hilbert space $\big(\mathcal{H},\langle\cdot,\cdot\rangle_{\scriptscriptstyle{H}}\big)$, 
extending  $(\cdot,\cdot)_{\scriptscriptstyle{H}}$ to $\mathcal{H}$.

Moreover, $\big(\mathcal{H},\langle\cdot,\cdot\rangle_{\scriptscriptstyle{H}}\big)$ has a \emph{countable orthonormal basis} $\big(\eta_k\big)_{k\in\mathbb{N}}$. For $k\in\mathbb{N}$ let $\lambda_k\in\mathbb{R}$ such that 
\begin{align*}
1<\lambda_1<\lambda_2<\ldots<\lambda_k<\lambda_{k+1}<\ldots\quad\text{and}\quad\sum_{k=1}^\infty\frac{1}{\lambda_k^2}<\infty.
\end{align*}
Next we consider
\begin{align*}
\mathcal{H}\ni f\mapsto Af:=\sum_{k=1}^\infty\lambda_k\,\big\langle f,\eta_k\big\rangle_{\scriptscriptstyle{H}}\,\eta_k\in\mathcal{H}
\end{align*}
and define for $p\in\mathbb{N}$
\begin{align*}
\mathcal{H}_p:=\big\{f\in\mathcal{H}\,\big|\,\Vert A^pf\Vert_{\scriptscriptstyle{H}}<\infty\big\}\quad\text{and}\quad \mathcal{N}:=\bigcap_{p\in\mathbb{N}}\mathcal{H}_p,
\end{align*}
where $\Vert\cdot\Vert_{\scriptscriptstyle{H}}$ denotes the induced norm on $\mathcal{H}$.    
Then $\mathcal{N}$ is a countably Hilbert space, which is Fr\'{e}chet and nuclear, compare e.g.~\cite{Ob94}. Its topological dual is given by 
$$
\mathcal{N}':=\bigcup_{p\in\mathbb{N}}\mathcal{H}_{-p}.
$$
Thus we obtain the Gel'fand triple
$$
\mathcal{N}\subset \mathcal{H} \subset\mathcal{N}'.
$$
We denote complexifications by a subscript $\C$.

Now by the Bochner-Minlos-Sazanov theorem, see e.g.~\cite{BK95} or \cite{hida70}, we define a Gaussian measure $\mu_{\scriptscriptstyle{H}}$ on $ \mathcal{N}'$ by 
$$\int_{\mathcal{N}'} \exp\left(i \langle \omega, \xi \rangle_{\mathcal{H}} \right)\, d\, \mu_{\scriptscriptstyle{H}}(\omega) := \exp\left(-\frac{1}{2} \| \xi \|_{\scriptscriptstyle{H}} \right).$$

\begin{remark}
Note that the measure has full support, i.e.~every open set has positive measure. This can be seen by \cite{KSW}, Thm.~6 or the 
fact that the measure is quasi translation invariant w.r.t.~shifts in direction of the subspace $\mathcal{N}$ dense in $\mathcal{N}'$, compare e.g.~\cite{HKPS93}, chapter 4B.

\end{remark}
We obtain the probability space $(\mathcal{N}', \mathcal{C}_{\sigma}, \mu_{\scriptscriptstyle{H}} )$.
Here $\mathcal{C}_{\sigma} :=\sigma(\mathcal{C}^{\xi_1, \dots , \xi_n }_{F_1, \dots , F_n})$ denotes the $\sigma$-algebra of cylinder sets
\begin{multline}
\mathcal{C}^{\xi_1, \dots , \xi_n }_{F_1, \dots , F_n} 
= \Big\{ \omega \in \mathcal{N'} \, \big| \langle \xi_1,\omega \rangle \in F_1, \dots ,\langle \xi_n, \omega \rangle \in F_n, \\
 \xi_i \in \mathcal{N}, F_j \in \mathcal{B}(\R), j=1,\dots ,n, \,\, n \in \N\Big\},
\end{multline}
where $\mathcal{B}(\R)$ denotes the $\sigma$-algebra of Borel sets in $\R$. 

Note that since $\mathcal{N}$ is a nuclear countably Hilbert space we have, see e.g.~\cite{HKPS93}:
$$
\mathcal{C}_{\sigma}(\mathcal{N}')=\mathcal{B}_w(\mathcal{N}') =\mathcal{B}_s(\mathcal{N}'),
$$
where $\mathcal{B}_w(\mathcal{N}')$ (resp.~$\mathcal{B}_s(\mathcal{N}')$) is the Borel $\sigma$-algebra generated by the weak (resp. strong) topology.

We define by 
$$\mathcal{P}:= \left\{ p\in L^2(\mathcal{N}'; \mu_{\scriptscriptstyle{H}}) \,\Big|\, p= \sum_{n=0}^N \langle \omega^{\otimes n }, f^{\otimes n } \rangle , \quad f \in \mathcal{N}_{\C}\right\}$$ 
the space of smooth polynomials.

We intend to construct the stochastic quantization Markov process via a local Dirichlet form as e.g.~in \cite{Fukushima}. To this end we define differential operators as follows: 
\begin{definition}
Let $p\in\mathcal{P}$
and $(\eta_k)_{k\in \N} \subset  \mathcal{N}$ a CONS of $\mathcal{H}$. 
Setting 
\begin{align*}
D_{\eta_k} p (\omega) = \lim_{\lambda \to 0} \frac{p(\omega + \lambda \eta_k) - p(\omega)}{\lambda}=\sum_{n=1}^N n \langle \eta_k \otimes \omega^{\otimes n-1} , f^{\otimes n} \rangle,
\end{align*}
we define 
$$\nabla p := (D_{\eta_k} p )_{k=1}^{\infty}.$$
\end{definition}
\begin{remark}
Note that this defines $D_{\eta_k}$ and $\nabla$ on a dense subspace of $L^2(\mathcal{N}'; \mu_{\scriptscriptstyle{H}})$.
\end{remark}

For $p\in\mathcal{P}$ we have
\begin{multline*}
\sum_{k=1}^{\infty} (D_{\eta_k} p)^2 = \sum_{k=1}^{\infty} \sum_{m=1}^N \sum_{n=1}^N m n \langle \omega^{\otimes m-1}\otimes \eta_k , f^{\otimes m} \rangle \langle \omega^{\otimes n-1}\otimes \eta_k , f^{\otimes n} \rangle \\
= \sum_{m=1}^N m \sum_{n=1}^N n \langle \omega^{\otimes m+n+2}, (f,f)_{\mathcal{H}} f^{\otimes n+m-2} \rangle.
\end{multline*}
Furthermore for $ u \in \mathcal{H}$ the adjoint 
$D^*_u  = \langle \cdot, u \rangle_{\scriptscriptstyle{H}} - D_u $ on a dense subspace, e.g.~polynomials in $L^2(\mathcal{N}'; \mu_{\scriptscriptstyle{H}})$, see e.g.~\cite{Po97}.

\section{Results}

In the following we will just write $L$ for $L(B^H)$, where $B^H$, $H\in (0,\frac{1}{d})$, is a $d$-dimensional fractional Brownian motion with Hurst parameter $H$. 

Moreover we denote by $\nu_g$ the measure $\nu_g = e^{-gL}\mu_H$.

\begin{theorem}\label{thm closable} 
The bilinear form
$$ \mathcal{E}_{\nu_g}(u,v) := \mathbb{E} (e^{-gL} \nabla u \cdot \nabla v), \quad u,v \in \mathcal{P},$$
is a densely defined, closable, symmetric pre-Dirichlet form and gives rise to a local, quasi-regular Dirichlet form $(\mathcal{E}_{\nu_g},D(\mathcal{E}_{\nu_g}))$.
Here $\mathbb{E}$ denotes expectation w.r.t.~$\mu_{\scriptscriptstyle{H}}$.  
\end{theorem}

\begin{theorem}\label{thm diffusion}
There exists a diffusion process $\mathbf{M} = (\Omega, \mathcal{F}, (\mathcal{F}_t)_{t\geq0}, (X_t)_{t\geq 0}, (P_{\omega})_{\omega \in \mathcal{N}'})$ which is associated with 
$(\mathcal{E}_{\nu_g}, \mathcal{D}(\mathcal{E}_{\nu_g}))$.
\end{theorem}

\begin{theorem}\label{thm irreducible}
	There exists a constant $c_0>0$ (see Lemma 4.1 below), such that for all $g<c_0$ the form
$(\mathcal{E}_{\nu_g}, D(\mathcal{E}_{\nu_g}))$ is irreducible (i.e. $u\in D(\mathcal{E}_{\nu_g}) \text{ with } (\mathcal{E}_{\nu_g}(u,u)=0$ implies $u$ is a constant), equivalently the associated diffusion is invariant under time translations.
\end{theorem}

\section{Proofs}
\noindent{\bf Proof of Thm.~\ref{thm closable}:} Note that the polynomials are dense in $L^2(\mathcal{N}';e^{-gL} \mu_{\scriptscriptstyle{H})}$, compare the proof of Prop.~10.3.~\cite{HKPS93}, p.371 and Prop.~2.3.2 from \cite{Ob94}.
We show closability using an integration by parts criterion. 
Let $(\eta_k)_k$ be a CONS. We can write 
$$\mathcal{E}_{\nu_g}(u,v) = \sum_{k=1}^{\infty} \mathbb{E} \left(e^{-gL} D_{\eta_k} u \cdot D_{\eta_k}v \right).$$
It is enough to show, see e.g.~\cite{MR92}, that each term in the sum is closable. 
Indeed we have for the components of the gradient:
\begin{multline}\label{eq:grad_comp}
\mathbb{E} (D_{\eta_k} u \cdot (v \exp(- gL))) = \mathbb{E}( u D_{\eta_k}^* (v \cdot \exp(-gL)))\\
=\mathbb{E}\left( u\left(\langle \cdot, \eta_k \rangle v \exp(-gL) - D_{\eta_k}v-vg D_{\eta_k}L\right)\exp(-gL)\right).
\end{multline}
We intend to show that the last expression is finite for $u,v \in \mathcal{P}$. This is evident for the first two terms. For the last one we have
$$
\left|\mathbb{E}\left(v (D_{\eta_k}L)\exp(-gL)\right)\right| \leq \|v\|_{L^2(\mathcal{N}'; \mu_H)} \cdot \| D_{\eta_k}L\|_{L^2(\mathcal{N}'; \mu_H)},
$$
which is finite due to the Meyer-Watanabe differentiability of $L$, see e.g.~\cite{Hu2001}.
For $u,v \in \mathcal{P}$ the expression \eqref{eq:grad_comp} is well-defined, i.e.~the adjoint of the gradient is densely defined.
Hence the form is closable, see e.g.~\cite{Kato76}.\\
The rest follows by \cite{MR92} Section IV.b., Thm. 3.5. 

 $\hfill \blacksquare$

\noindent{\bf Proof of Thm.~\ref{thm diffusion}}
The proof is an immediate result of \cite{MR92} Section IV.b., Thm. 3.5.  
 $\hfill \blacksquare$\\
For the proof of Theorem~\ref{thm irreducible} we will follow the lines of \cite{ARHZ}. In order to do so, we need a further result from \cite{HNS06}.

\begin{lemma}[\cite{HNS06}, Thm. 1]
Suppose that $Hd<1$. Then the self-intersection local time $L$ fulfills
$$\mathbb{E}(e^{L^p})< \infty,$$
for any $p<\frac{1}{Hd}$. Moreover there exists a constant $c_0>0$ such that for all $c < c_0$ one has
$$\mathbb{E}(e^{c L^{\frac{1}{Hd}}})< \infty,$$
\end{lemma}  
In particular, $\mathbb{E}(e^{c L})$ is then finite. This we will use in the following proof. \\

\noindent{\bf Proof of Thm.~\ref{thm irreducible}:}  
For $\varphi\in L^2(\mathcal{N'},e^{-gL} \mu_{\scriptscriptstyle{H}} )$, and $\frac{1}{p}+\frac{1}{q}=1, p,q>1$ we have
\begin{multline*}
\int |\varphi|^{\frac{2}{p}} \, d\mu_{\scriptscriptstyle{H}} (\omega)  = \int |\varphi|^{\frac{2}{p}} \exp\Big(-\frac{g}{p} L\Big) \exp\Big( \frac{g}{p}  L\Big)\, d\mu_{\scriptscriptstyle{H}} (\omega)\\ 
\leq  \left(\int |\varphi|^{2} \exp(-g L) \, d\mu_{\scriptscriptstyle{H}} (\omega)\right)^\frac{1}{p} \cdot  \left(\int \exp\Big( \frac{q}{p} g L\Big) \, d\mu_{\scriptscriptstyle{H}} (\omega)\right)^{\frac{1}{q}},
\end{multline*}
by H\"older inequality.
The last term is finite if $\frac{q}{p} g< c_0$ as in Lemma 4.1. In this case we obtain 
\begin{align}\label{eq:Hoelder}
\int |\varphi|^{\frac{2}{p}} \, d\mu_{\scriptscriptstyle{H}} (\omega) \leq  C \|\varphi\|^{\frac{2}{p}}_{L^2(\mathcal{H};e^{-gL}d\mu_{\scriptscriptstyle{H}} )}.
\end{align}
Hence for any polynomial $v\in\mathcal{P}$, by setting $\varphi = \nabla v$, we have
\begin{equation}\label{eq:grad}
\|\nabla v\|^2_{\scriptscriptstyle{H,\frac{2}{p}}}:=\left(\int \|\nabla v \|_{\scriptscriptstyle{H}}^{\frac{2}{p}} d\mu_{\scriptscriptstyle{H}} (\omega) \right)^{\frac{2}{p}} \leq C\, \mathcal{E}_{\nu_g}(v,v).
\end{equation}
So we see already that $\mathcal{E}_{\nu_g}(v,v)=0$ implies $v=const.$ a.e.~for any $v\in\mathcal{P}$.

Consider now $u \in  D(\mathcal{E}_{\nu_g})$ such that $\mathcal{E}_{\nu_g}(u,u)=0$. Then by closability we find a sequence $(v_n)_n$ in $\mathcal{P}$ such that $v_n \to u$ in $D(\mathcal{E}_{\nu_g})$. 
Then by \eqref{eq:grad} we see that $\|\nabla v_n \|_{\scriptscriptstyle{H, \frac{2}{p}}} \to 0$. 
 The convergence in $L^2(\mathcal{N}'; e^{-gL} d\mu_{\scriptscriptstyle{H}})$ implies convergence in $L^{\frac{2}{p}}(\mathcal{N}', d\mu_{\scriptscriptstyle{H}})$, due to \eqref{eq:Hoelder}. The same holds for the gradients by \eqref{eq:grad}.
 
Altogether, we have a polynomial sequence which approximates $u$ in $D_{\scriptscriptstyle{1, \frac{2}{p}}}$. Here $D_{\scriptscriptstyle{1, \frac{2}{p}}}$ denotes the corresponding Malliavin--Sobolev space, see e.g.~\cite{Nualart} Sect.~1.2. Thus by \cite{Nualart} Prop.~1.5.5.~we obtain $u \in D_{\scriptscriptstyle{1, \frac{2}{p}}}$, if $\frac{2}{p}>1$, which is the case if we choose $p\in (1,2)$.

Moreover, in this case we have that $\nabla u=0$ a.e..
Then by \cite{Nualart} Prop.~1.2.5 , since $\nabla u =0$ and $u \in D_{\scriptscriptstyle{1, \frac{2}{p}}}$, we find that $u = \mathbb{E}(u),$ which is a constant.

The properties hold $\mu_{\scriptscriptstyle{H}}$-a.e. and thus $e^{-gL} \mu_{\scriptscriptstyle{H}}$-a.e.~by absolute continuity of the measure. 

Thus we have shown the assertion. $\hfill \blacksquare$

\noindent{\bf{Acknowledgement:} }We truly thank M.~R\"ockner for helpful discussions. Furthermore we thank M.~Grothaus and M.~J.~Oliveira for helpful comments. Financial support by CRC 701 and the mathematics department of the University of Kaiserslautern for research visits at Bielefeld university are gratefully acknowledged.

\end{document}